\documentclass[11pt,a4paper]{article}
\usepackage[hyperref]{emnlp2020}
\usepackage{times}
\usepackage{latexsym}

\usepackage{microtype}

\usepackage{float}
\usepackage[utf8]{inputenc} 
\usepackage[T1]{fontenc}    
\usepackage{hyperref}       
\usepackage{url}            
\usepackage{booktabs,multirow}       
\usepackage{amsfonts}       
\usepackage{nicefrac}       
\usepackage{microtype}      

\usepackage{graphicx}
\usepackage{times}
\usepackage{epsfig}
\usepackage{graphicx}
\usepackage{amsmath}
\usepackage{amssymb}
\usepackage{xcolor,vwcol}

\DeclareMathAlphabet{\pazocal}{OMS}{zplm}{m}{n} 
\usepackage{mathtools}
\DeclarePairedDelimiterX{\norm}[1]{\lVert}{\rVert}{#1}
\usepackage{multicol}
\usepackage[linesnumbered,ruled,vlined]{algorithm2e}
\usepackage{setspace}

\let\oldnl\nl
\newcommand{\nonl}{\renewcommand{\nl}{\let\nl\oldnl}}

\DeclareMathAlphabet{\pazocal}{OMS}{zplm}{m}{n} 
\usepackage{microtype}
\usepackage{graphicx}
\usepackage{subfigure}
\usepackage{booktabs} 
\usepackage{times}
\usepackage{latexsym}
\usepackage{makecell} 
\usepackage{amsfonts} 
\usepackage{amsmath} 
\usepackage{amssymb} 

\aclfinalcopy 
\def\aclpaperid{1435} 

\title{RecoBERT: A Catalog Language Model for Text-Based Recommendations}

\author{Itzik Malkiel\textsuperscript{1,2}
,Oren Barkan\textsuperscript{1,3}
,Avi Caciularu\textsuperscript{1,4}
,Noam Razin\textsuperscript{1,2}
,Ori Katz\textsuperscript{1,5}
  \textbf{and} \textbf{Noam Koenigstein}\textsuperscript{1,2}\\ 
  \textsuperscript{1}Microsoft\\
  \textsuperscript{2}Tel Aviv University\\
  \textsuperscript{3}Ariel University\\
  \textsuperscript{4}Bar-Ilan University\\
  \textsuperscript{5}Technion \\
  {\tt \small \{itmalkie, orenb, Ori.Katz, Noam.Koenigstein\}@microsoft.com}\\
  {\tt \small avi.c33@gmail.com}
  \\
  {\tt \small noam.razin@cs.tau.ac.il}
  }

\date{}

\begin{document}

\maketitle

\begin{abstract}
Language models that utilize extensive self-supervised pre-training from unlabeled text, have recently shown to significantly advance the state-of-the-art performance in a variety of language understanding tasks. However, it is yet unclear if and how these recent models can be harnessed for conducting text-based recommendations. In this work, we introduce RecoBERT, a BERT-based approach for learning catalog-specialized language models for text-based item recommendations. We suggest novel training and inference procedures for scoring similarities between pairs of items, that don't require item similarity labels. Both the training and the inference techniques were designed to utilize the unlabeled structure of textual catalogs, and minimize the discrepancy between them. By incorporating four scores during inference, RecoBERT can infer text-based item-to-item similarities more accurately than other techniques.
In addition, we introduce a new language understanding task for wine recommendations using similarities based on professional wine reviews. As an additional contribution, we publish annotated recommendations dataset crafted by human wine experts. Finally, we evaluate RecoBERT and compare it to various state-of-the-art NLP models on wine and fashion recommendations tasks. 
\end{abstract}

\section{Introduction}
Recommendation systems are a major component of content discovery in online stores. Different recommendation systems are employed across a broad spectrum of domains, such as movies, music, groceries, and more. In each case, the recommendation system is associated with a different catalog of items comprising different descriptors, item properties, and metadata. 
This work deals with the case of generating item-to-item 
similarities based on item descriptions.

Personalized recommender systems make use of either or both Collaborative Filtering (CF) or Content-Based (CB) information \cite{aggarwal2016recommender}. CF approaches build models based on users past behavior \cite{breese2013empirical, schafer2007collaborative}. On the other hand, CB recommenders use item meta-data such as properties, tags, and descriptions in order to build and match user and item profiles \cite{brusilovski2007adaptive,wang2018content,lops2011content}. A model that utilizes both CF and CB is called a hybrid recommender system.

Item-to-item recommendations are commonly used in large scale recommender systems such as on Netflix \cite{gomez2015netflix}, Amazon \cite{linden2003amazon}, Xbox \cite{XboxMovies} and many others. Commonly found on product details page (PDP), these non-personalized recommendation lists are known to drive-up purchases as well as user engagement. Similar to personalized recommendation, item similarities can be computed based on user activity, item meta-data or both, using a variety of different models. 
In a new store, where user data does not exist, item-to-item recommendations are computed using one or more content-based approaches that leverage item meta-data in order to compute item-to-item similarities. The extracted data may include images, videos, textual descriptions, and more.

Textual content-based recommendation systems leverage textual information about items, such as item descriptions and titles. These models usually rely on Natural Language Processing (NLP) models to compute item-to-item similarities. A naive approach to produce recommendations from textual information is to infer similarities by embedding the textual description (and title) of every item in a latent space \cite{lops2011content,wang2018content,de2015semantics}. Item embeddings, that utilize textual descriptions, can be obtained via different types of language models. 

Recently, self-supervised pre-training of language models have revolutionized the field of NLP. These techniques first utilize a self-supervised pre-training of a neural-based model using a large corpus of unlabeled text. Then, apply fine-tuning for specific NLP tasks. 
Among the recent self-supervised pre-trained language models, BERT \cite{devlin2018bert} has emerged as a very powerful method, achieving state-of-the-art results in a variety of NLP tasks such as sentiment analysis \cite{sun2019utilizing}, language inference \cite{wu2019beto,cui2019pre}, sentence similarities \cite{reimers2019sentence} and more. BERT pre-training technique incorporates (1) reconstruction of randomly masked words (known as masked language model), and (2) predicting whether two sentences are consecutive (next sentence prediction).

In this work, we build upon BERT and introduce a novel technique for self-supervised pre-training of catalog-based language models. In addition, we introduce an inference technique that utilizes the above model for inferring item similarities that can be used for item-to-item recommendations in cold catalogs. 
Hence, we name our technique RecoBERT - a BERT model adapted for textual based recommendations.

RecoBERT pre-training leverages self-supervision to its fullest by utilizing a combination of a masked language model along with a title-description model. The latter comprises a learning task that reveals relationships between item titles and descriptions. In some cases, these relations can form a summarization task, for which titles are short sentences that summarize the longer descriptions. In other cases, catalogs may comprise items with implicit titles that incorporate a few words that were crafted for each item at hand. For both cases, the title-description task encourages the model to reveal the underlying connections between titles and descriptions, improves language understanding, and therefore yields more accurate embeddings. This results in an improved text-based item similarity performance in cold catalogs. Importantly, RecoBERT doesn't require item similarity labels nor usage data.

We also introduce a new NLP wine recommendation task, demonstrating RecoBERT's ability to find similar items in very complex domains. The task utilizes a publicly available dataset comprising ~120K elaborate wine descriptions written by wine experts. The goal is to produce wine recommendations for each item in the dataset, in the form of other similar wines. We employed a professional wine sommelier to manually craft 1095 recommendations for $\sim$100 wines that form a ``ground-truth'' test-set for evaluations. For reproducibility, and as an additional contribution, we made these annotations publicly available\footnote{https://doi.org/10.5281/zenodo.3653403}.

Importantly, the novel wine recommendations task introduced in this work is different and more complex than most NLP tasks usually considered. The wine reviews incorporate domain-specific semantics, taxonomy, and phrases, as well as picturesque descriptions of tastes, aromas, and colors. Arguably, determining similarities between wine reviews is a challenging task, which requires a high level of intelligence and knowledge even to the average human. Specifically, compared to the tasks presented in the GLUE benchmark \cite{wang2018glue}, for which the average adult person can easily solve a query in few seconds, determining the similarity of wines based on their reviews may pose a challenge to most people and takes up to a few minutes even to wine enthusiasts and professionals. 

The main contribution of this paper is threefold: (1) We introduce RecoBERT, a self-supervised training for catalog-based language model. (2) We introduce a novel inference technique that yields item-to-item similarities by leveraging RecoBERT, and compare its performance against relevant baselines. (3) we introduce a novel complex NLP task of wine recommendations and publish a matched labeled test set crafted by a professional sommelier.

\begin{figure*}[t]
\includegraphics[width=1.0\linewidth]{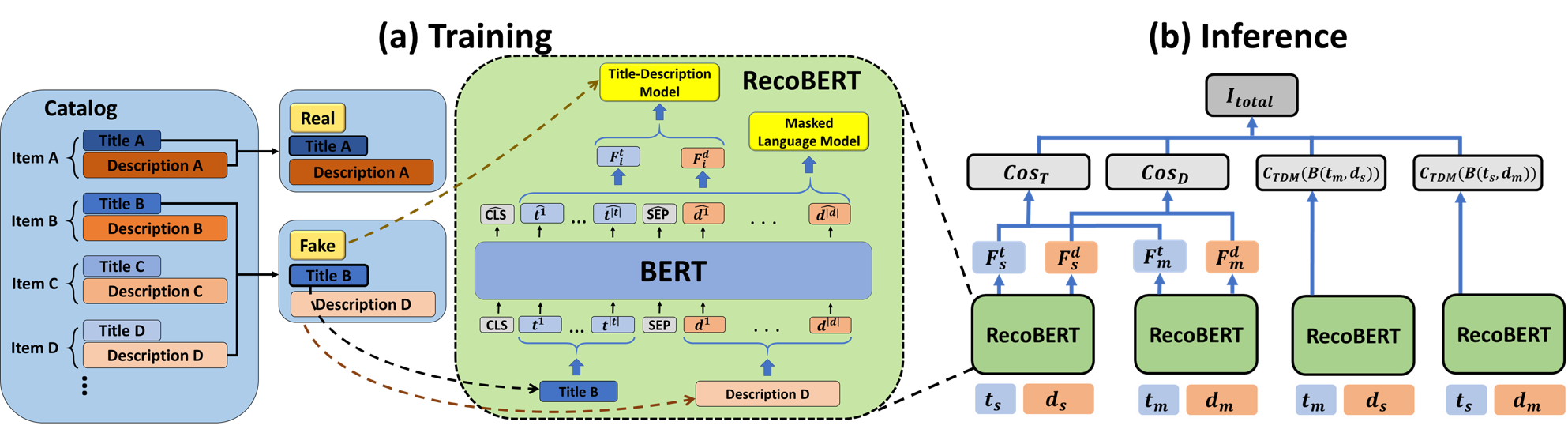}
  \centering
  \caption{RecoBERT receives title-description pairs corresponding to \emph{positive} (``real'') and \emph{negative} (``fake'')  samples, extracted from a given catalog. (a) during training, the title-description pairs are propagated through the BERT backbone and transformed into two feature vectors. These vectors are then fed into the TDM, minimizing a cosine loss between them. 
  (b) in inference, four scores are computed. Two scores propagate the seed and candidate items separately (``real'' pairs). The other two scores utilize the TDM head and propagate title-description pairs extracted from both seed and candidate items (``fake'' pairs). 
  }
    \label{fig:architecture}
\end{figure*}

\section{Related Work}

Recent methods in text-based recommendations suggest a hybrid approach that combines usage data with either traditional or neural-based NLP methods.
In \cite{reconn,zheng2017joint}, the authors suggest a hybrid approach for recommendations that utilizes both session data (CF) and textual features from articles extracted by a convolutional neural network (CNN). Additionally, in \citet{wang2015collaborative,djuric2015hierarchical} the authors proposed hierarchical Bayesian models for learning a joint representation for textual content and personal ratings, using latent Dirichlet allocation (LDA), deep autoencoders, and word2vec \cite{mikolov2013distributed}. In contrast to the above methods, the model in this paper doesn't depend on usage data and hence can be applied to completely cold catalogs.
Recently, \cite{hashtag} proposed attentive CNN for performing hashtag recommendations for tweets. This method solely depends on text, but requires supervision for similarity. Unlike this method, our model focuses on textual catalogs and doesn't require item-to-item similarity labels.

A recently proposed family of Transformer-based language models \cite{devlin,liu2019roberta,yang2019xlnet} uses multiple attention layers and a two-phase training procedure composed of unlabeled pre-training and supervised fine-tuning. These models show great promise in linguistic tasks, and were shown to exceed human-level baselines in specific tasks such as machine translation \cite{NIPS2017_7181}, question answering \cite{yang2019xlnet}, 
and other related tasks \cite{wang-etal-2018-glue}. These models utilize \textit{sentence embedding} techniques \cite{palangi2016deep}, where a text is encoded into low dimensional vectors that summarize the information in the input text. For example, in universal sentence encoder (USE) \cite{cer2018universal}, the authors suggest utilizing vectors extracted from a machine translation model for transfer learning to other NLP tasks. 

Lately, \citet{wang2019superglue,storks2019commonsense,assenmacher2020comparability} claimed that human baselines are being surpassed by Transformer-based models and others that exploit statistical cues in the well-known GLUE set \cite{wang2018glue}. Such models may suffer severe performance degradation when putting to use on real-world problems. Hence, some argue that the tasks in the GLUE dataset no longer suffice for evaluating language understanding models.

In this work, we propose a new language task that is much more complicated than the semantic similarity tasks in GLUE. Motivated by extracting item similarities for recommender systems, our task is neither composed of single sentences nor sentence pairs. Instead, the goal is to induce semantic similarity between wine items represented by sentence-paragraph pairs. Due to the complexity of the wines domain, as well as the professional language and length of the wine reviews, our novel language understanding task requires a high level of intelligence and knowledge that exceeds the average human level.

\section{Methodology}

Let $\mathcal{W}=\left\{w_{i}\right\}_{i=1}^{W}$ be the vocabulary of tokens in a given language. Let $T$ be the set of all possible sentences generated by $\mathcal{W}$, including the empty sentence. Additionally, let $D$ be the set of all possible paragraphs generated by $T$. Let $\mathcal{C} := \{m_i|m_i \in T \times D\}_{i=1}^{c}$ be a catalog of items, where each item is associated with a title-description pair (titles are sentences, and descriptions are paragraphs). 
Given a catalog $\mathcal{C}$, the task is to infer a similarity function $\mathcal{F} :  \mathcal{C} \times \mathcal{C}  \rightarrow  \mathbb{R}$, that scores the similarity between any pair of items $s,m \in \mathcal{C}$. In particular, $\mathcal{F}$ can be used to quantify a similarity score that ranks all the items in the catalog according to their semantic textual similarity with a given seed item $s \in \mathcal{C}$. 

\subsection{Model Architecture and Loss Functions}

RecoBERT is a function $\mathcal{B}: T \times D \rightarrow \mathbb{R}^{h} \times \mathbb{R}^{h} $, which utilizes a BERT-Large architecture \cite{devlin} with a hidden layer size of $h$, and incorporates (1) a title-description model (TDM) for scoring the relation between titles and descriptions, and (2) a mask language model (MLM) for specializing in a given domain. A dataset of $n$ training samples is represented as pairs ${(t_i, d_i) \in T \times D}$, indexed by $i = 1...n$. Each pair is associated with a label $y_i \in \{0,1\}$, indicating whether $t_i$ and $d_i$ correspond to the same item. 

Following the MLM procedure in \cite{devlin}, RecoBERT transforms each $(t_i, d_i)$ pair into sequences of inputs tokens $(t^j_i)_{j=1}^{|t|}$ and $(d^j_i)_{j=1}^{|d|}$, masks 15\% of them and adds the special $CLS$ and $SEP$ tokens. 
This input sequence is then mapped to a sequence of latent embedding tokens by propagating the input through BERT
{\label{BERT}
\begin{multline*}
BERT(I(t_i, d_i)) := \\ (\widehat{CLS}_i, (\widehat{t_i^j})_{j=1}^{|t|}, \widehat{SEP}_i, (\widehat{d_i^j})_{j=1}^{|d|})
\end{multline*}}
where each latent token corresponds to its matched input token. 

Two feature vectors are then computed by ${F_i^t := \frac{1}{|t|}\sum_{j=1}^{|t|} \widehat{t_i^j}}$ and ${F_i^d := \frac{1}{|d|}\sum_{j=1}^{|d|} \widehat{d_i^j}}$. Importantly, $F_i^t$ and $F_i^d$ correspond to the title and description of the input, respectively. 

It is important to clarify the distinction between BERT and RecoBERT. Bert yields contextualized embeddings (as defined in Eq.~\ref{BERT}), and can be replaced by any other language model. On the other hand, RecoBERT is defined as:
\begin{equation}
\label{RecoBERT}
\mathcal{B}(t_i,d_i) = (F_i^t, F_i^d),
\end{equation}

RecoBERT loss function is composed of two components, a TDM loss, and an MLM loss. The purpose of TDM is to learn the relationship between item titles and descriptions. To this end, we feed the model with both \emph{positive} (``real'') title-description pairs, for which both title and description belong to the same item, and \emph{negative} (``fake'') pairs, where the title and description are taken from two different items.

The TDM loss term utilizes a cosine head ${C_{TDM}: \mathbb{R}^{h} \times \mathbb{R}^{h} \rightarrow \mathbb{R}}$, that scores the relation between a title-description pair. Formally, 
\begin{equation}
{C_{TDM}(F_i^t, F_i^d) = \frac{1+\mathrm{cosine}(F_i^t, F_i^d)}{2}},
\end{equation}
and the TDM loss is defined as
\begin{multline*}
\pazocal{L}_{TDM} = -\frac{1}{n}\Sigma_{i=1}^{n}[y_i\log(C_{TDM}(F_i^t, F_i^d)) \\ +(1-y_i)\log(1-C_{TDM}(F_i^t, F_i^d))].
\end{multline*}

The purpose of the MLM is to specialize RecoBERT's language model on the specifics of the domain and catalog at hand. As we shall see later, this has major significance in complex NLP tasks such as wine recommendations where the semantic meaning of certain words differs from their usual semantic meaning.  

The MLM loss follows the paradigm presented in \cite{devlin2018bert}, utilizes a classifier $C_{MLM}: \mathbb{R}^{d} \rightarrow \mathbb{R}^{|\mathcal{W}|}$ that projects the embedded tokens to the vocabulary space, and applies a softmax function to infer pseudo-probabilities. The MLM loss function can be expressed as 
${\pazocal{L}_{MLM} = - \frac{1}{n} \Sigma_{i=1}^{n}\Sigma_{(l,k) \in z_i} \log{(C_{MLM}(e_l)_k)}}$, 
where $z_i$ is a sequence of index pairs $(l,k)$ that correspond to the $i$th training sample, $l$ and $k$ are the indices of the masked token in $BERT(I(t_i,d_i))$ and the vocabulary $\mathcal{W}$, respectively. In summary, the total loss for RecoBERT is defined as ${\pazocal{L}_{total} = \pazocal{L}_{MLM} + \pazocal{L}_{TDM}}$. 

\subsection{Training}
\label{Model training}
We split the dataset into a train and validation sets. The validation set is used for early stopping, as we have found it essential, especially for smaller-sized datasets. RecoBERT backbone is initialized by the prescribed weights of the publicly available pre-trained BERT model, while the TDM head is initialized from scratch. 

During training, we iterate over the items in the train set, generating \emph{positive} and \emph{negative} samples by switching the description to that of another item with probability $p_s=0.5$.
Then, the \emph{positive}  and \emph{negative} labels are assigned accordingly.  
The RecoBERT model and training is illustrated in Fig.~\ref{fig:architecture}(a).

\subsection{Inference}
\label{section:inference}
RecoBert's inference proceeds by generating four scores. First, we propagate every item $(t_i,d_i) \in C$ through RecoBERT, extracting $F_i^t$ and $F_i^d$, as defined in Eq.~\ref{RecoBERT}. Then, given a seed item $s = (t_s, d_s) \in  \mathcal{C}$, and for any item $m \neq s, m = (t_m,d_m) \in \mathcal{C}$, we define the two cosine scores denoted by $Cos_D(s, m) := \mathrm{cosine}(F_{s}^d, F_{m}^d)$ and  $Cos_T(s, m) := \mathrm{cosine}(F_{s}^t, F_{m}^t)$. These two cosine scores represent the similarity between (1) the seed and candidate titles, and (2) the seed and candidate descriptions.

Next, we utilize the learned TDM head to compute additional two cosine scores. Specifically, we propagate the pairs $(t_m, d_s)$ and $(t_s, d_m)$ through RecoBERT, extracting $C_{TDM}{({B}(t_m, d_s))}$ and $C_{TDM}({B}(t_s, d_m))$, respectively. These two scores approximate the similarity between the candidate title and the seed description, and between the seed title and the candidate description.

Finally, we normalize each score separately, across all candidate items, to have a zero-mean and a unit-variance, and define the total score as follows:
\begin{multline}
\pazocal{I}_{total}(s, m) = \lambda_1 Cos_D(s, m) + \lambda_2 Cos_T(s, m) + \\ \lambda_3  C_{TDM}{({\mathcal{B}}(t_m, d_s))} + \lambda_4  C_{TDM}({\mathcal{B}}(t_s, d_m)),
\end{multline}
where $\lambda_1 \dots \lambda_4$ are set to $1$, and the item-to-item recommendations are obtained by sorting the candidate items according to $\pazocal{I}_{total}$, in a descending order. RecoBert's inference scheme is depicted in Fig.~\ref{fig:architecture}(b).

\section{Wine Recommendations from Reviews}
We introduce a novel NLP recommendation task of finding wine recommendations from reviews. 
The task is based on a publicly available dataset from Kaggle\footnote{https://www.kaggle.com/zynicide/wine-reviews}, and a new test set, annotated by a professional wine sommelier. 
A common obstacle in evaluating similarity models is the lack of a relevant test-set or ground-truth.  Therefore, as part of this paper's contributions, we made this test publicly available. 
The Kaggle dataset, together with our annotated ground truth, form a new text-based recommendation task that can be further used by others in the future. 

\subsection{The Wine Dataset}

The Kaggle wine dataset comprises of 120K wine titles and reviews. Each title is composed of: (1) winery name, (2) wine year, (3) wine name, and (4) grape variety. The reviews are single paragraphs descriptions written by wine experts, delineating taste, aromas, and other wine characteristics. 

The descriptions frequently use a nonliteral, symbolic jargon common with wine enthusiastic and Oenologists. For example, wine sweetness can be identified by five intensity levels, including bone-dry, dry, off-dry, sweet, and very sweet. These intensity levels substantially affect the similarity between wines. 
Hence, the task of wine recommendations might be considered as more complex and more difficult than many other classical NLP tasks such as sentiment analysis or question answering. While these classical tasks are relatively very simple for most humans, the wine recommendations task is arguably more difficult and convoluted even to intelligent humans.

Generally, inferring wine similarity requires the solution of the following language understanding challenges:

\paragraph{1. Characteristic Intensities} Wines comprise different characteristics with different intensity levels. 

\paragraph{2. Characteristic Categories}
Taste and aroma are classified into associative categories, and some classes are more distinct than others. For example, \emph{apple} and \emph{citrus} are two distinct categories of taste. Given a wine with a \emph{hint of apple}, a recommendation for a wine with \emph{citrus characteristics} is inadvisable by most professionals. In this example, the additional difficulty stems from the fact that a general (non-specialized) language model may consider ``apple'' and ``citrus'' to be relatively close as both are fruits.

\paragraph{3. Domain-specific Semantics and Taxonomy}
Compared to general language, the wine domain incorporates professional jargon with unique phrases, different semantics, and unique taxonomy. 
For example, the semantic opposite of the word \emph{dry} in the English language is usually the word \emph{wet}, however, in the context of wines, it is the word \emph{sweet}. Similarly, the opposite of \emph{white} is generally \emph{black} where in the wine domain it is the word \emph{red}. 

\paragraph{4. Non-literal Figurative Descriptions} 
Professional wine reviews incorporate symbolic descriptions that depart from their literal meaning. For example, one reviewer unfavorably described a wine named "Riscal 1860" using the words "Bulky and clumsy", which implies that the combination between acidity, tannins, alcohol, and sugars, is out of balance. 

\begin{figure}[t]
\includegraphics[width=1.0\linewidth]{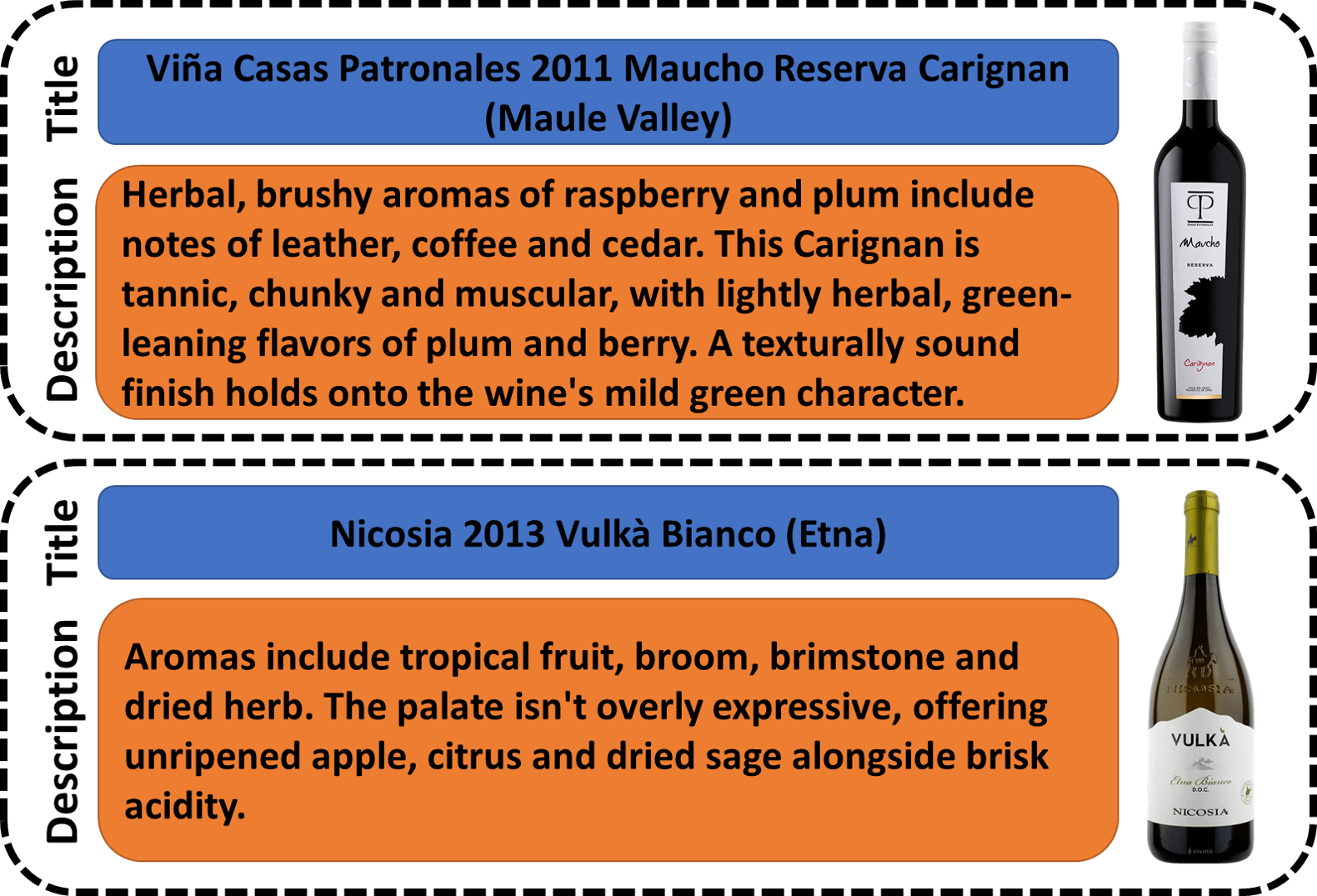}
  \centering
  \caption{Two items from the wine dataset, each composed of a title-description pair. Images are shown for illustration.}
    \label{fig:wine_dataset}
\end{figure}

Fig.~\ref{fig:wine_dataset} presents two representative samples from the dataset. The top example is a red wine, named ``Maucho Reserva''. Its description incorporates domain-specific phrases, such as ``tannic'', categorial flavors, such as ``raspberry'', ``plum'', ``coffee'' and more. The description incorporates figurative terms, such as ``chunky and muscular'' and ``texturally sound finish''. The second example is ``Vulk\'a Bianco'', a white with a relatively straight forward description expressing the different flavor categories, and the intensity level of the ``acidity''.

\begin{figure*}[t]
\includegraphics[width=1.0\linewidth]{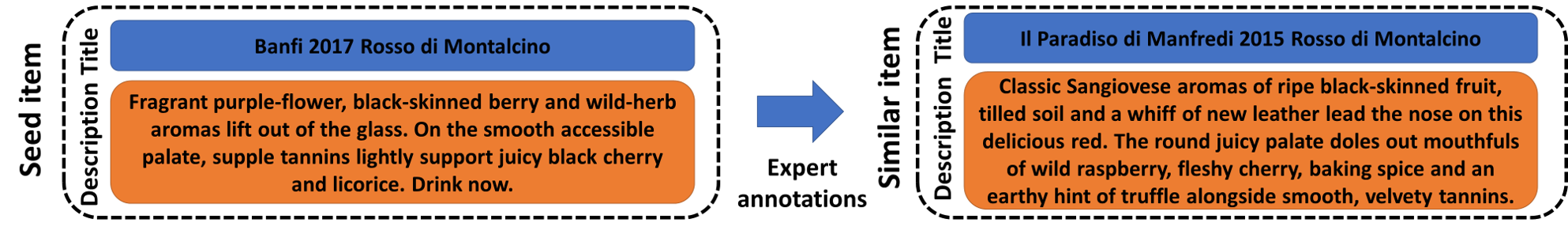}
  \centering
  \caption{A representative sample from our annotated wine recommendations dataset, crafted by human wine experts. 
  }
    \label{fig:expert_annotation}
\end{figure*}

\subsection{The Expert Annotations Set}

Unlike collaborative filtering models, Content-based item to item similarity/recommendation models are very hard to evaluate. Hence, we collected a test set, annotated by professional wine sommeliers, comprising of 1095 wine recommendations to 100 wines. The sommeliers were asked to choose representative ``seed'' items and annotate each with $\sim$10 other wines that share similar characteristics with the seed item. For the sake of reproducibility and as an additional contribution, we made these annotations publicly available\footnote{https://doi.org/10.5281/zenodo.3653403}. 

Fig~\ref{fig:expert_annotation} exhibits one sample from the annotated expert recommendations. As can be seen in the figure, the seed and the recommended item share similar phrases, such as ``ripe of black-skinned fruit'', ``smooth'' and ``velvety tannins'' from the recommended item, that can be associated with phrases in the seed item, including ``black-skinned berry'', ``smooth accessible palate'' and ``supple tannins''.

\section{Evaluation}
We evaluate RecoBERT on two datasets: (1) wines catalog and (2) fashion catalog. For each catalog, we train a separate RecoBERT model 
using the Adam optimizer \cite{kingma2014adam} and a batch size of 16.

\subsection{Baseline Models}

We compare RecoBERT with the following models:
 
\paragraph{Universal Sentence Encoder} (USE) suggests to leverage feature vectors extracted from a Transformer model \cite{vaswani2017attention} for transfer learning tasks. The Transformer architecture is composed of encoder and decoder networks. During the forward pass, the Transformer receives text in a source language, forwards it through the encoder, outputs a feature vector, feeds it into the decoder, which then generates text in the target language. USE \cite{cer2018universal} utilizes the above intermediate feature vector for transfer learning to other NLP tasks, including semantic textual similarity (such as STS Benchmark \cite{agirre2012semeval}), sentiment analysis \cite{sun2019utilizing}, etc. In our work, we employ USE to generate separate embedding for every item title and description.

\paragraph{Pre-trained-BERT} is the pre-trained BERT-Large model from \cite{devlin2018bert}. This model was trained using a large corpus of unlabeled text, to both optimize the masked language model and the next sentence prediction (NSP) task. Since, in most datasets, item similarity labels do not exist, we can not fine-tune this model for the item similarity task. Instead, we utilize the pre-trained BERT model as a feature extractor, and extract the feature vectors $F^t_m$ and $F^d_m$ (see Equ.~\ref{RecoBERT}), for every item in the catalog.

\paragraph{Specialist-BERT} is a BERT-Large model that continued pre-training using a domain-specific corpus. Specifically, we create a specialized corpus by extracting the description paragraphs of all the items in the given catalog. Then, we iterate over sentence pairs extracted from the above corpus and continue training the pre-trained BERT with the identical BERT pre-training technique, as presented at \cite{devlin2018bert}. We train this model with similar settings used for RecoBERT (i.e. train-validation split, 1.5M training steps, etc.). Feature vectors are extracted in the same way as for the above Pre-trained BERT model.

\paragraph{MoverScore} employs a contextualized embedding model and a variant of the Earth Mover Distance \cite{rubner2000earth} to measure the similarity between sentence-pairs \cite{moverscore}. Given two sentences, MoverScore aligns similar words from each sentence and computes the flow traveling between these words. MoverScore has recently emerged as a promising text similarity metric for text generation tasks, including summarization, machine translation, image captioning, and data-to-text generation. In our experiments, we utilize the MoverScore technique on top of the Specialist-BERT model.

\paragraph{Inferencing} baseline models, besides the MoverScore, utilize the inference technique presented in section  \ref{section:inference}, by setting $\lambda_3$ and $\lambda_4$ to $0$ (i.e. applying the sum $\lambda_1 Cos_D(s, m) + \lambda_2 Cos_T(s, m)$). For the USE inferencing, we replace the underlying feature vectors $(F_{s}^d, F_{m}^d)$ and $(F_{s}^t, F_{m}^t)$ by those extracted from USE.
The MoverScore baseline is applied with its own scoring technique\cite{moverscore}, utilizing the EMD between the latent representations of the words.

\subsection{Quantitative Metrics}
\paragraph{Hit Ratio at $k$ (HR@$k$)} 
HR@$k$ is the percentage of the predictions made by the model, where the true item was found in the top $k$ items suggested by the model. Specifically, a seed-candidate pair is scored with $1$ if the candidate item is ranked within the top $k$ recommendations produced by the model w.r.t. to the seed, otherwise $0$. Then the average over all seed-candidate examples in the test set is reported.
\paragraph{Mean Reciprocal Rank (MRR)}
This measure is defined as the average of the reciprocal ranks considering the entire set of ranked items (and not just the top-$k$). In contrast to HR, the MRR metric takes into consideration the exact order within the recommendation list.
\paragraph{Mean Percentile Rank (MPR)}
Given a seed item, the percentile rank is the rank that was assigned by the recommendation model to the correct item (to be retrieved), divided by the number of ranked items. This quantity computed for all the items in the test set and then being averaged.

For more details, we refer the reader to \cite{resnick1997recommender}.

\begin{table*}[t]
\centering
\smallskip\noindent
\resizebox{0.9\linewidth}{!}{%
\begin{tabular}{cclccccccc} 

& & \thead{Model} & \thead{MPR} & \thead{MRR} &\thead{HR@1000}  & \thead{HR@100}  & \thead{HR@50}  & \thead{HR@10}  & \thead{HR@5} \\
\hline

\parbox[t]{2mm}{\multirow{10}{*}{\rotatebox[origin=c]{90}{\textbf{120K wines}}}}&

\parbox[t]{2mm}{\multirow{5}{*}{\rotatebox[origin=c]{90}{\textbf{descriptions}}}}

&USE$_{\lambda_2\gets0}$    &   77.1\% & 4.3\% &  12.2\%  &  4.1\% & 2.9\% &   0.7\% & 0.5\%     \\

&&Pre-trained-BERT$_{\lambda_2\gets0}$ &   80.8\%  & 6.9\%& 21.7\% & 6.5\% & 4.0\% & 1.2\% & 1.0\%  \\

&&Specialist-BERT$_{\lambda_2\gets0}$ &   96.2\% & 11.6\% &  44.5\%& 17.2\%& 11.1\%& 4.5\% & 3.2\%  \\

&&MoverScore$_{descriptions}$ &   96.8\% & 11.45\% &  46.84\%& 19.84\%& 12.34\%& 5.93\% & 4.81\%  \\

&&RecoBERT$_{\lambda_2,\lambda_3,\lambda_4\gets0,0,0}$   &   \textbf{97.3\%}  &  \textbf{21.0\%} &   \textbf{64.9\%} &  \textbf{33.2\%} &  \textbf{24.9\%} &
\textbf{10.4\%}& \textbf{6.6\%}  \\

\cmidrule{2-10}

&\parbox[t]{2mm}{\multirow{5}{*}{\rotatebox[origin=c]{90}{\textbf{titles}}}}

&USE$_{\lambda_1\gets0}$    &   83.2\% & 5.6\% &  14.3\%  &  6.5\% & 3.5\% &   0.9\% & 0.6\%     \\

&&Pre-trained-BERT$_{\lambda_1\gets0}$     &   85.6\%  & 8.9\%& 23.7\% & 8.5\% & 6.2\% & 2.1\% & 1.3\%  \\

&&Specialist-BERT$_{\lambda_1\gets0}$&   96.6\% & 13.7\% &  46.6\%& 19.4\%& 13.6\%& 5.26\% & 3.5\%  \\

&&MoverScore$_{both}$ &   95.6\% & 14.56\% &  48.24\%& 21.25\%& 14.79\%& 6.86\% & 3.75\%  \\

&&RecoBERT$_{\lambda_1, \lambda_3, \lambda_4\gets0,0,0}$   &   \textbf{97.8\%}  &  \textbf{23.5\%} &   \textbf{68.4\%} &  \textbf{35.6\%} &  \textbf{27.8\%} &
\textbf{12.5\%}& \textbf{8.9\%}  \\

\midrule

\parbox[t]{2mm}{\multirow{5}{*}{\rotatebox[origin=c]{90}{\textbf{Expert subset}}}}&

\parbox[t]{2mm}{\multirow{5}{*}{\rotatebox[origin=c]{90}{\textbf{titles \& desc.}}}}

&USE    &   72.4\% & 18.3\% &  97.3\%  &  34.4\% & 21.8\% & 8.2\% &  4.7\%     \\

&&Pre-trained BERT           &   76.8\%  & 24.3\%& 97.7\% & 44.5\% & 32.8\% & 12.3\% & 6.6\%    \\

&&Specialist-BERT &   92.3\% & 35.1\% &  \textbf{99.9}\% & 79.6\%& 59.3\%& 25.0\% & 14.7\%   \\

&&MoverScore$_{both}$ &   93.5\% & 54.4\% &  99.8\%& 80.2\%& 67.8\%& 35.8\% & 20.7\%  \\

&&RecoBERT$_{\lambda_3, \lambda_4\gets0,0}$   &   95.2\%  & 90.3 \% &   99.9\% &  84.2\% &  72.0\% & 60.6\% & 23.0\%   \\

&&RecoBERT&   \textbf{96.3\%}  &  \textbf{91.7\%} &   99.8\% &  \textbf{94.9\%} &  \textbf{89.6\%} & \textbf{65.4\%} & \textbf{38.6\%} \\
\hline
\end{tabular}}
\caption{ Recommendations results evaluated on the 120K wines dataset (upper part), and the subset of 1095 annotated items (bottom part). 
}
\label{Tab:quant_baseline_inference}
\end{table*}

\begin{table}[t]
\centering
\smallskip\noindent
\begin{tabular}{lcc}
\thead{Model}  &\thead{Average rank}  \\
\hline

Pre-trained-BERT & 3.21 \\

USE & 3.58 \\
Specialist-BERT & 3.60 \\
MoverScore$_{both}$ &   3.75   \\

RecoBERT     &  \textbf{3.94}    \\
\hline
\end{tabular}
\caption{Expert evaluation for fashion recommendations.}
\label{Tab:fashion}
\end{table}

\subsection{Wine Recommendations Results}
For the wine dataset, we compare RecoBERT with all four baselines by three different evaluations. The first two evaluations conduct item similarities by solely relying on item descriptions or item titles (but not both), and ranking 120K wine items. The third evaluation utilizes both item titles and descriptions, ranking the subset of the expert annotated wines.

In Tab.\ref{Tab:quant_baseline_inference}, we report MPR, MRR, and five HR@k scores, for each evaluation, using the 1095 expert annotations. 
In the upper and middle parts of the table, all models solely utilize item descriptions and item titles, respectively. In both evaluations, each model ranked the entire 120K wines in the catalog, for each seed. To make a clean comparison between RecoBERT and the other BERT-based models, in these experiments, we have evaluated all BERT-based models (including RecoBERT) with the same inference score. Specifically, for the descriptions evaluations (upper part) we set all BERT-based models to solely use the $Cos_D(s, m)$ score (by configuring $\lambda_1$ to $1$ and setting the other $\lambda$s with $0$, i.e. we set $\lambda_2$, $\lambda_3$ and $\lambda_4$ in RecoBERT to $0$, and $\lambda_2$ to $0$ in the other BERT-based baselines). In similar, the titles evaluations (middle part) we solely utilize the $Cos_T(s, m)$ score (setting  $\lambda_2$ to $1$ and eliminating the rest of the scores). The MoverScore in each section utilizes the textual information associated with its name. 

In the bottom part of the table, we report the performance of all models, utilizing both item titles and descriptions, comparing against the full RecoBERT inference, as presented in the section \ref{section:inference}. In these evaluations, the reported MoverScore${_{both}}$ separately applies the MoverScore on item titles and descriptions, ranking the items in the catalog by computing the sum of both scores. 

The results in the table indicate that RecoBERT outperforms all other models, in all three categories. Specifically, by solely utilizing item descriptions, RecoBERT results with MPR of 97.3\% while the baselines models yield an MPR of 77.1\% (USE),  80.8\% (Pre-trained-BERT), 96.8\% (MoverScore), and 96.2\% (specialist-BERT). For MRR, RecoBERT scored 21.04\%, while the baseline models ranged between 4.31\% (for USE) and 11.6\% (for specialist-BERT). In addition, RecoBERT presents superior performance on all HR metrics, sometimes improving by a factor of two, even compared to specialist-BERT and MoverScore which yield the best performance among the baseline models. This can be attributed to the importance of the title-description learning task, and to the benefit gained by the TDM head, which produces more accurate embedding under a cosine metric. 

Notably, in the same description-based evaluations (upper part of the table), RecoBERT yields 10.4\% in the HR@10 metric. This entails that on average, for each seed, RecoBERT was able to retrieve roughly one out of $\sim$10 expert annotations, in the top ten recommendations list, by ranking 120K candidate items. Remarkably, $\sim$10 annotated items represent $\sim$0.0083\% of the entire catalog.  

Additionally, as can be seen in the bottom part of the table, RecoBERT with the full inference yields better performance, by a sizeable margin, compared to all other models, including the same RecoBERT applied with the baseline inference (which solely utilizes the $Cos_D(s, m)$ and $Cos_T(s, m)$ scores). The latter is evidence for the benefit of applying the full inference method, which also utilizes the TDM head by propagating title-description pairs extracted from seed-candidate items.

\subsection{Fashion Recommendations Results}

We evaluate RecoBERT on a fashion catalog incorporating 4K items and compare its performance with all four baseline models. Similar to the wines evaluations, all BERT-based models were initialized with the Pre-trained BERT weights and continued pre-training using the text extracted from the fashion catalog. During inference, all models used both item titles and descriptions.

To assess the quality of the recommendations, we report human scoring conducted by a fashion expert. The same test set, composed of 100 seed items, was ranked by all models. The scoring was performed blindly, as the source model for each sample was hidden from the expert. For each seed, the expert ranked the top five recommended items, by a total score of 0 to 5, indicating poor to excellent performance 

As can be seen in Tab.~\ref{Tab:fashion}, RecoBERT outperforms all baselines, including the ones that utilize the BERT model that was specialized in the fashion domain. Specifically, RecoBERT has gained a relative improvement of 9.4\% and 5.9\% compared to specialist-BERT and MoverScore, respectively. See the supplementary materials for more results of RecoBERT applied to the fashion dataset.

\subsection{Ablation Study}
\label{ablation}

\begin{table}[t]
\centering
\smallskip\noindent
\begin{tabular}{lccc} 
\thead{Model} & \thead{MPR} & \thead{MRR} & \thead{HR@10} \\
\hline

RecoBERT$_{\lambda_3, \lambda_4\gets0,0}$   &   95.2\%  &  90.3\% &   60.6\%   \\

RecoBERT$_{\lambda_1, \lambda_2\gets0,0}$  &  88.5\%  &  75.5\% &   45.8\%   \\

RecoBERT$_{\lambda_1\gets0}$   &   92.6\%  &  80.5\% &   50.1\%   \\ 
RecoBERT$_{\lambda_2\gets0}$   &   88.7\%  &  77.2\% &   48.5\%   \\ 
RecoBERT$_{\lambda_3\gets0}$   &   95.5\%  &  90.6\% &   60.8\%   \\ 
RecoBERT$_{\lambda_4\gets0}$   &   95.3\%  &  90.8\% &   62.3\%   \\ %

RecoBERT&   \textbf{96.3\%}  &  \textbf{91.7\%} & \textbf{65.4\%} \\
\hline
\end{tabular}
\caption{ Ablation study results}
\label{Tab:ablation}
\end{table}

Tab.~\ref{Tab:ablation} presents an ablation study for RecoBERT inference, evaluated on the subset of the wine expert annotations. Six variants are considered, each eliminates different scores from RecoBERT inference, by setting their matched $\lambda$s with $0$. The results, shown in the table, indicate that it is crucial to employ all four scores, in the way it is done in RecoBERT, and that extracting information from both item titles and descriptions is highly beneficial for item similarity performance.

{\color{black} \subsection{Computational Costs}
We report computation times that were measured for RecoBERT training and inference, by utilizing a single NVIDIA V100 32GB GPU using PyTorch framework. For the wines catalog, we trained RecoBERT for 1.5M training steps. This training took $\sim$5 days. RecoBERT training on the fashion catalog, comprised 150K steps and took $\sim$12 hours. Inferencing RecoBERT with the same GPU allows a throughput of ~340 items per second. This enables recommending the entire fashion catalog in $\sim$7 hours,
and the wines test set in ~9.5 hours. 
Notably, all these computations are applied once, for a given catalog, and can be executed in an offline manner and cached for later use. To further accelerate the computation time of the two $C_{TDM}$ scores applied through RecoBERT inference, one can adopt knowledge distillation techniques, such as \cite{barkan2019scalable, jiao2019tinybert,lioutas2019distilled}, which are beyond the scope of this work.

}
\section{Discussion and Conclusions}

In this work, we introduce a novel natural language recommendation task along with a novel annotated test set that together contribute to the state-of-the-art research of text-based recommenders and language models. We present RecoBERT - A model for text-based item similarity that (a) mitigates the discrepancy between training and inference phases in the classical BERT model, by operating on sentence-paragraph pairs, (b) refines the backbone language model to provide more accurate embeddings, improving item similarities under the cosine metric, and (c) utilizes matched cosine scores as part of the inference process. In addition, we show that the unique mechanism behind RecoBERT leads to significant improvements over the other baselines and across all metrics.

RecoBERT's preeminence stems from two properties of its TDM loss: First, feeding title-description pairs allows RecoBERT to apply cross-attention between the tokens of both elements entailing an effective dependency between their embeddings. Second, by leveraging the TDM task, RecoBERT learns an additional task for revealing the underlying connections between item titles and descriptions, which reinforces the model to better specialize in the domain at hand. 

In some cases, where titles do not correlate with item descriptions\footnote{E.g. in the case of movies, titles are more cryptic and less informative.}, RecoBERT can be extended to utilize textual tags\footnote{For example, in movies, tags can include actors, genre, etc.}, by either concatenating the tags with the title, or feeding them as an extra input to RecoBERT.

Compared to other semantic textual similarity tasks, the proposed wine recommendations task, along with our published annotated test set, can shed light on the limitations as well as the key advances of state-of-the-art NLP models for recommendations. In addition, by publishing our annotated wine recommendations dataset, we intend to encourage the community to further explore the boundaries of other NLP models, assessing the ability 
of machines to understand complex human language.

\section*{Acknowledgment}
As part of this work, we have productized a similar capability in Microsoft, which utilizes the TNLR language model \cite{tnlrv3} as a backbone. Under this context, we would like to thank Saurabh Tiwary and Saksham Singhal for many helpful discussions and advice.

\bibliographystyle{acl_natbib}
\bibliography{example_paper}

\clearpage

\twocolumn[{%
\begin{center}
{\LARGE Appendix}
\end{center}
\appendix

\begin{center}
\includegraphics[width=1\linewidth]{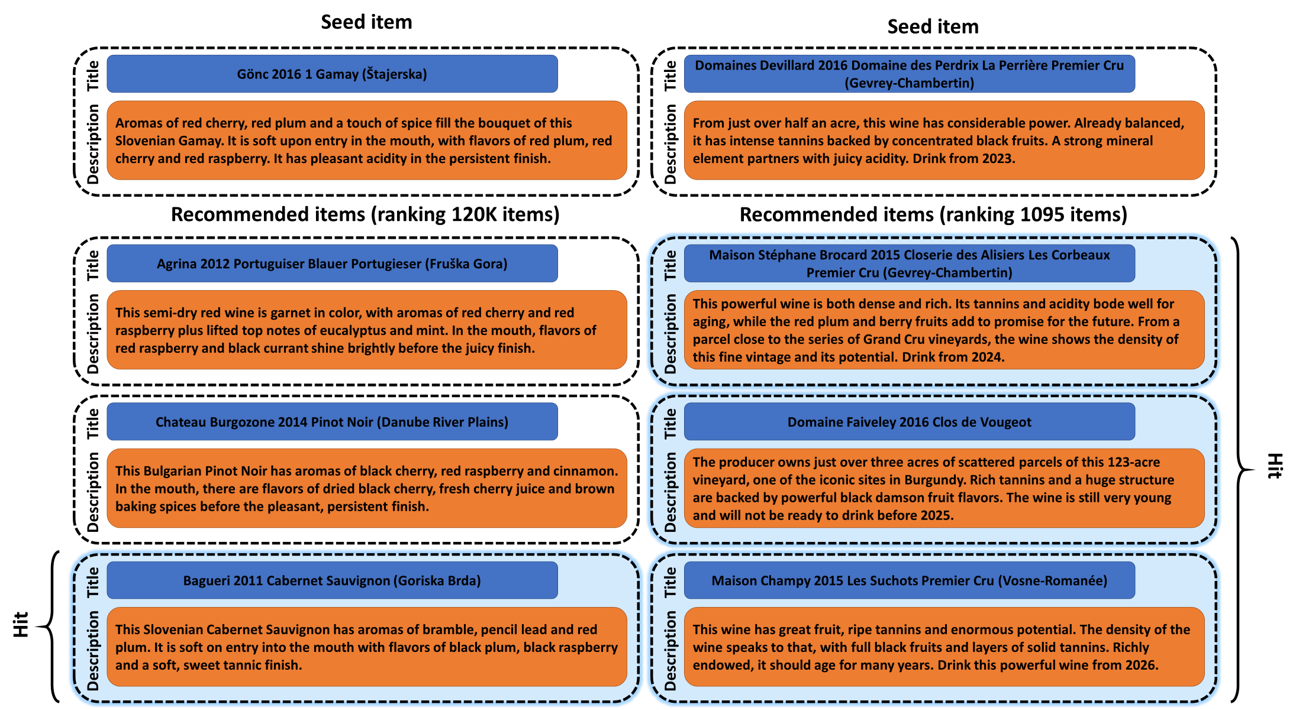}
\captionof{figure}{Two representative seeds along with their top three recommended items retrieved by RecoBERT. The left column presents RecoBERT's top three recommendations while ranking the 120K dataset. The right column presents the recommendations for ranking the 1095 expert annotated items. Highlighted recommended items were annotated by the human expert as similar to the seed (``hit'').}
\label{fig:quali_wines2}
\end{center}
}]

\begin{figure}[t]
\includegraphics[width=0.9\linewidth]{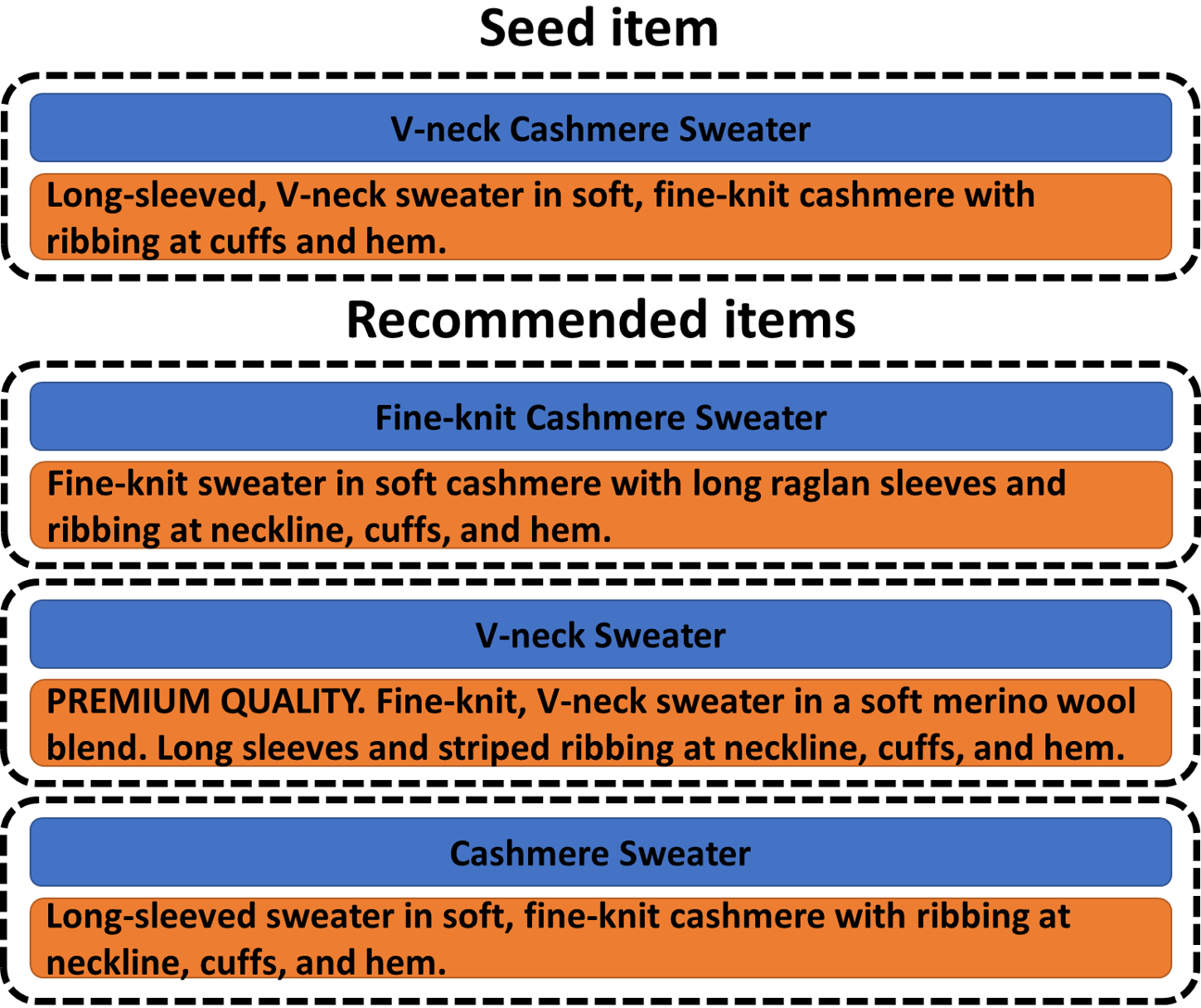}
  \centering
  \caption{Fashion results. A representative seed item associated with the top three recommended items retrieved by RecoBERT. 
  }
    \label{fig:quali_fashion}
\end{figure}

\section{Qualitative Results}

Fig~\ref{fig:quali_wines2} exhibits the top three recommendations for two representative wine seeds from the annotated test set, predicted by RecoBERT, while (1) ranking the entire 120K wines in the catalog (left column), and (2) ranking the subset of 1095 annotated dataset (right column). As can be seen in the figure, the third item retrieved by RecoBERT was also annotated by the human wine expert (``hit''). Specifically, for the ``Gamay" seed, the origin of the first recommended wine is a neighbor country of the seed's origin, and both items share similar berries aromas and body intensity. The second wine is a Pinot Noir, known as Gamay's ``cousin", and shares similar berries aromas. The third item, which is one out of ten expert annotations for this seed, shares the same seed origin, aromas, and flavors. 

For the seed ``Domaines Devillard'', the top three recommendations retrieved by RecoBERT were all annotated by the wine expert. For this seed, all the recommendations share the same aging potential. Also, both the seed, first and third items are Cru wines, sharing the same origin and variety. The origin of the second item is Burgundy, which is considered a high-quality vineyard, similar to Cru.


Fig.~\ref{fig:quali_fashion} exhibits a representative seed from the fashion catalog, along with the top three recommendations ranked by RecoBERT. As can be seen in the figure, the top three recommended items for "V-neck Cashmere Sweater" are (1) consistent with the sweaters sub-category, (2) comprise the same material or indicated by a "PREMIUM QUALITY" label, and (3) preserve similar style properties, such as V-neck, and ribbing in a few locations on the sweater.

\end{document}